\documentclass[12pt]{article}
\usepackage[pdftex]{color,graphicx}
\usepackage{mathabx}
\usepackage{ifsym}
\usepackage{floatflt}
\usepackage{wrapfig}
\usepackage{eufrak,txfonts}

\def\ca{{\cal A}}
\def\cb{{\cal B}}

\def\d{\delta}
\def\e{\epsilon}

\def\g{\gamma}

\def\m{\mu}
\def\n{\nu}

\def\t{\tau}

\def\G{\Gamma}

\def\S{\Sigma}

\def\ca{{\cal A}}
\def\cb{{\cal B}}

\def\cf{{\cal F}}

\def\ch{{\cal H}}

\def\cm{{\cal M}}

\newcommand{\One}{{\boldmath 1}}

\begin{document}

\vskip 15mm

\begin{center}

{\Large\bfseries Holonomy Loops, Spectral Triples
\\[2mm]
\&  Quantum Gravity$^\ast$ \\[2mm]  
}

\vskip 4ex

Johannes \textsc{Aastrup}$\,^{a}$\footnote{email: \texttt{johannes.aastrup@uni-muenster.de}},
Jesper M\o ller \textsc{Grimstrup}\,$^{b}$\footnote{email: \texttt{grimstrup@nbi.dk}}\\ \& Ryszard \textsc{Nest}\,$^{c}$\footnote{email: \texttt{rnest@math.ku.dk}}

\vskip 3ex  

$^{a}\,$\textit{SFB 478 "Geometrische Strukturen in der Mathematik"\\
  Hittorfstr. 27, D-48149 M\"unster, Germany}
\\[3ex]
$^{b}\,$\textit{The Niels Bohr Institute \\Blegdamsvej 17, DK-2100 Copenhagen, Denmark}
\\[3ex]
$^{c}$ \textit{Matematisk Institut\\ Universitetsparken 5, DK-2100 Copenhagen, Denmark}
\end{center}

\vskip 2ex

\begin{abstract}
We review the motivation, construction and physical interpretation of a semi-finite spectral triple obtained through a rearrangement of central elements of loop quantum gravity. The triple is based on a countable set of oriented graphs and the algebra consists of generalized holonomy loops in this set. The Dirac type operator resembles a global functional derivation operator and the interaction between the algebra of holonomy loops and the Dirac type operator reproduces the structure of a quantized Poisson bracket of general relativity. Finally we give a heuristic argument as to how a natural candidate for a quantized Hamiltonian might emerge from this spectral triple construction.\\
\vspace{1.6cm}
\end{abstract}

\noindent$^\ast${\small based on talk given by J.M.G. at the QG2 conference, Nottingham, juli 2008; at the QSTNG conference in Rome in sept/oct 2008; at the AONCG conference, Canberra, december  2008.}

\newpage

\section{Introduction}

Noncommutative geometry \cite{ConnesBook} has proven a remarkable successful framework for understanding the geometrical nature of the standard model of particle physics. 
The pioneering work of Alain Connes and co-workers on the standard model grounds on the observation that the Dirac operator on a compact manifold together with its interaction with the smooth functions on the manifold and its representation on square integrable spinors completely characterizes the Riemannian structure. This leads to the notion of a {\it spectral triple} $(\ca,\ch,D)$, where $\ch$ is a separable Hilbert space, $\ca$ is a commutative involutive algebra represented in $\ch$ and $D$ is a self-adjoint operator in $\ch$ and where $\ca$ and $D$ has to satisfy certain conditions.  
Moreover, these conditions 
are naturally extended to encompass also noncommutative algebras. This generalization turns out to be the right framework to formulate the standard model {coupled} to general relativity as a single gravitational theory \cite{Connes:1996gi,Chamseddine:1991qh,Chamseddine:1996rw,Chamseddine:1996zu,Chamseddine:2006ep,Chamseddine:2007hz,Chamseddine:2007ia}. In particular, the classical action of the standard model coupled to gravity emerges from a spectral action principle applied to an algebra which is a noncommutative modification of the algebra of smooth functions over $M$ \cite{Chamseddine:1991qh,Chamseddine:1996rw,Chamseddine:1996zu}. In this formulation the additional interactions (strong, electro-weak) emerge through inner fluctuations of $D$ generated by the noncommutativity of the algebra.

This beautiful formulation of fundamental physics raises the question whether quantum field theory, too, has a natural translation into the language of noncommutative geometry. The standard model, after all, is a quantum field theory. When formulated in terms of noncommutative geometry one recovers, however, essentially its classical formulation. Moreover, the action of the standard model emerges as an integrated part of a gravitational theory. Therefore, it appears plausible that {\it if} a natural intersection between quantum field theory and noncommutative geometry exist, then it should involve aspects of quantum gravity.

This line of reasoning is the motivation behind a programme \cite{Aastrup:2005yk,Aastrup:2006ib,Aastrup:2008wa,Aastrup:2008wb,Aastrup:2008zk}  which we initiated in 2005 and which will be the topic here. Basically, the idea is to adopt a top-down approach to quantum gravity by applying elements of noncommutative geometry to a quantum gravity setup. In particular, we take inspiration from loop quantum gravity \cite{Thiemann:2001yy,Rovelli:2004tv,Ashtekar:2004eh}, which is based on Ashtekars formulation of general relativity as a gauge field theory. However, whereas loop quantum gravity  is based on an algebra of Wilson loops of the Ashtekar gauge fields, the idea here is to consider instead the algebra of holonomy loops. Since this algebra is intrinsically noncommutative we are immediately situated well within the domain of noncommutative geometry. To complete the analogy we construct a Dirac type operator and require its interaction with this algebra of holonomies to reproduce the quantized Poisson structure of general relativity. Since holonomy loops are functions over connections this Dirac type operator will resemble a functional derivation operator over a space of connections.

The aim of this project is twofold. First, we aim to identify canonical structures and principles which may help to define a theory of quantum gravity at a quantized level and thereby avoid the ambiguities which plague the standard quantization procedure.
A Dirac type operator represents exactly such a canonical structure. The hope is that this operator and its interaction with the algebra of holonomy loops will cast light on the dynamics of quantum gravity. Indeed, in the last part of this paper we present a heuristic argument as to how a candidate for the Hamiltonian of quantum gravity may emerge from this spectral triple construction.

Second, the presence of inner automorphisms -- due to noncommutativity of the algebra -- will in general introduce an additional gauge sector. In Connes formulation of the standard model the entire bosonic sector arises through such inner automorphisms. Since the algebra of holonomy loops is noncommutative this mechanism will also apply to the spectral triple construction discussed here. Thus, although this construction grounds on a purely gravitational setup -- Ashtekars formulation of general relativity -- it will be a framework of quantum gravity which simultaneous includes a basic mechanism of unification. Put differently, it seems plausible that the construction {\it cannot} be a framework for pure quantum gravity. Clearly, we would like to interpret this "extra" in terms of matter degrees of freedom. Whether such an interpretation is valid can only be determined through a semi-classical analysis. 

The spectral triple construction involves several characteristics of quantum field theory. For instance, the spectral action functional, which is finite, resembles a Feynman integral over a space of connections. 
Also, the construction of the Dirac type operator naturally entails the {\it canonical commutations relations} (CAR) algebra, a hint at fermionic quantum field theory. The significance of these observations remains to be clarified.\\

 The note is organized as follows: in section \ref{LQG} we first give a brief introduction to loop quantum gravity with emphasis on topics which we will use in the following. In section \ref{PRO1} we present the general idea of the programme and in section \ref{PRO2} we provide the basic details of the spectral triple construction. Sections \ref{SPCO} to \ref{HAM} are concerned with the physical interpretation of the construction.

\section{Loop quantum gravity and spaces of connections}
\label{LQG}

Loop quantum gravity is based on Ashtekars formulation of general relativity in terms of gauge field variables \cite{Ashtekar:1986yd,Ashtekar:1987gu}. 
Upon selecting a foliation of the four dimensional manifold $M=R \times \Sigma$, where $\Sigma$ is a three dimensional hypersurface, the Ashtekar variables are given by
a $SU(2)$-connection $A^i_\n(x)$ and the inverse, densitized dreibein $E^\m_j(x)$, both fields on the hypersurface $\Sigma$. The connection is related to the extrinsic curvature of $\Sigma$ in $M$ and the dreibein is given by the intrinsic geometry of $\Sigma$. These are conjugate variables and 
satisfy the Poisson structure
\begin{equation}
\{ A_\n^i(x), E_j^\m (y)\}=\delta_\n^\m \delta^i_j\delta^3 (x-y).
\label{Poisson}
\end{equation}
These variables are subjected to three constraints, namely the Gau\ss-, Diffeo-morphism- and Hamilton Constraints. In one version the Hamilton constraint is of the form\footnote{In the following we shall ignore issues regarding the signature of space-time and the reality of the Irmirzi parameter which we set equal to one. Also, for simplicity we choose here a constant lapse field.} 
\begin{equation}
\int dx^3\epsilon^{ij}_{\;\;k}F^k_{\mu \nu}(x)E_j^\mu(x) E_i^\nu(x)\;,
\label{HAM}
\end{equation}
where $F$ is the field strength tensor of $A$. 

A key step in loop quantum gravity is the shift of variables from the set $(A,E)$ to a dual set of variables given by holonomies of $A$ and fluxes of $E$. Therefore, consider curves $\g$ in $\Sigma$ and the holonomy transforms
$$
h_\g(A )=\hbox{Hol} (\g,A )
$$
of $A$ along $\g$, and surfaces $S$ in $\Sigma$ and the fluxes

\begin{wrapfigure}{r}{40mm}
  \begin{center}
    \includegraphics{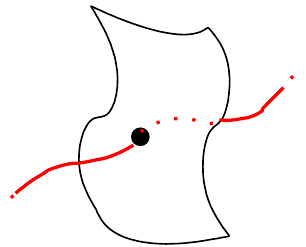}
  \end{center}
  \label{fig1}
\end{wrapfigure}
$$
F_S^a (E)=\int_S \epsilon^m_{\;\;np}E^{a}_mdx^n  dx^p 
$$
of $E$ over $S$. The Poisson bracket between these new variables read
\begin{equation}
\{ F_S^a(E),h_\g(A )\} =\pm h_{\g_1}(A) \tau^a h_{\g_2}(A)\;,
\label{bracket}
\end{equation}
where $\g = \g_1 \cdot \g_2$ intersects $S$ in the point $\g_1\cap \g_2$ (see figure) and where the sign corresponds to different orientation of the surface $S$ relative to $\g$.

This change of variables is crucial since it permits a reduction of the quantization procedure to a projective system of finite problems related to oriented graphs. Specifically, consider first a finite number of holonomy transforms along connected curves $\{\g_i\}$ in $\Sigma$. The union of these curves is a graph $\G$. We call the intersection points of the curves for vertices $\{v_j\}$ in $\G$ and the curves between vertices for edges $\{\e_k\}$. If we denote by $\ca$ the space of smooth $SU(2)$ connections on $\Sigma$ and consider the restriction of $\ca$ to $\G$, in terms of holonomies of connections along curves, then we see that $\ca$ reduces to the space 
$$
\ca_\G := G^{n(\G)}\;,
$$
where $n(\G)$ is the number of edges in $\G$. Here each copy of $G$ corresponds to the holonomy transform along an edge in $\G$.

Consider now the full system of {\it all} such piece-wise analytic graphs and their associated spaces $G^{n(\G_i)}$ 
\begin{figure}[h]
\begin{center}
\resizebox{!}{2.5cm}{%
 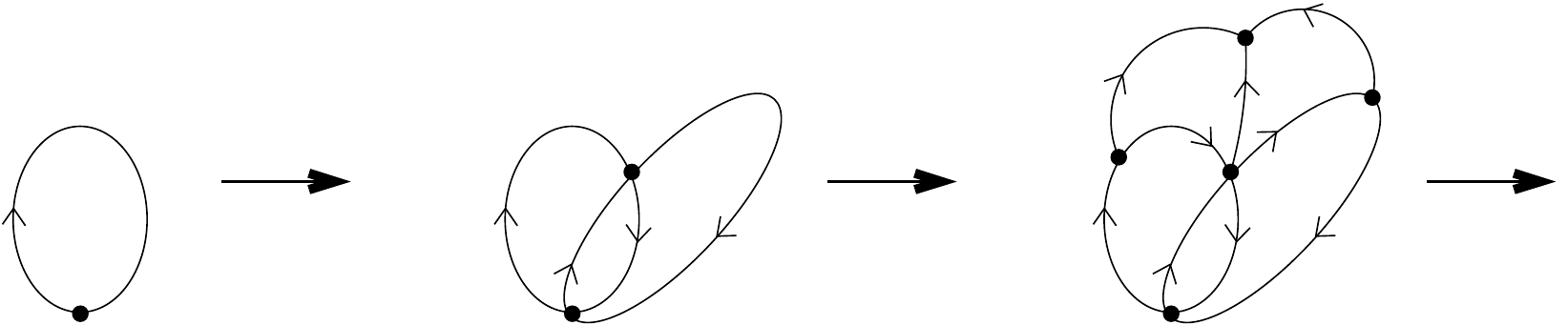}
\end{center}
\end{figure}

\noindent Given two graphs $\G_1\subset\G_2$ there exist projections between their associated spaces 
$$
P_{\G_2;\G_1}:\ca_{\G_2} \rightarrow \ca_{\G_1}\;.
$$ 
For example, take the two first graphs in the previous figure. The corresponding spaces are given by $\ca_{\G_1}=G$ and $\ca_{\G_2}=G^4$ and the projection between them reads (with labeling of edges from left to right)
$$
P_{\G_2;\G_1} (g_1,g_2,g_3,g_4) = g_1 \cdot g_3\;.
$$ 
A key result by Ashtekar and Lewandowski \cite{Ashtekar:1993wf,Ashtekar:1994wa} is that the space $\ca$ of smooth connections is densely embedded in the projective limit of spaces $\{\ca_\G\}$, that is
\begin{equation}
\ca\hookrightarrow \lim_{\leftarrow}\ca_\Gamma =:\overline{\ca}^{\small a}\;.
\label{LQGresult}
\end{equation}
Thus $\overline{\ca}^{\small a}$ can be considered as a completion of the space of connections. The main advantage of $\overline{\ca}^{\small a}$ is that it possesses a natural measure, the Ashtekar-Lewandowski measure. This measure is simply the inductive limit of Haar measures over spaces $\ca_\G$.   The Hilbert space $L^2(\overline{\ca}^{\small a})$ is the inductive limit of Hilbert spaces $L^2(\ca_\G)$. This is the kinematical Hilbert space of loop quantum gravity, denoted $H_{kin}$, and forms the basis of the theory. The quantized holonomy and flux operators $\hat{h}_L$ and $\hat{F_S}$ can be constructed as multiplication and derivation operators in $H_{kin}$ and one can impose the constraints as operator constraints and thereby, in principle, attempt to obtain the physical Hilbert space by solving these constraints.

It is important to notice that $H_{kin}$ is non-separable. This fact is the first indication that a spectral construction involving this Hilbert space will be hard to achieve.

\section{The project; idea and strategy}
\label{PRO1}

The aim of our project is to construct a spectral triple over an algebra of holonomy loops. This algebra is intrinsically noncommutative and the hope is that this noncommutativity will entail additional structure much alike the structures generated by the matrix factor in Connes' formulation of the standard model.

With an algebra of holonomy transforms, which we view as maps 
$$
\g: \nabla \rightarrow Hol(\g,\nabla)\in M_l(C)\;,
$$
where $l$ is the size of a matrix representation of $G$ and where $\nabla\in \ca$, it is clear that a spectral triple involving such an algebra will be a geometrical construction {\it over} the space of connection. That is, the space of states on the algebra will be related to the space $\ca$. Accordingly, the Dirac type operator in such a triple must have an interpretation in terms of functional derivations. 

The strategy to obtain such a construction is to exploit the pro-manifold structure of ${\ca}$. That is, since $\ca$ is densely embedded in $\overline{\ca}^{\small a}$ which is the projective limit of manifolds we aim to construct first geometrical structures over these intermediate manifolds and subsequently to take their projective and inductive limits over a suitable system of graphs. Thus, the programme involves the following three steps:
\begin{itemize}
\item[\# 1:]
construct spectral triples $(B_{\G},D_{\G},H_{\G})$ at the level of a finite graphs $\G$, with $B_\G$ being the algebra of holonomy loops, now generated by loops restricted to the graph $\G$. This step is conceptually easy since $\ca_\G $ is a compact Lie-group with a Haar measure which gives us a natural Hilbert space and candidates for a Dirac operator;
\item[\# 2:]
ensure compatibility with the structure maps 
\[
P_{\G\G'}: \ca_{\G}\rightarrow \ca_{\G'}\;,
\]
and their enduced maps between Hilbert spaces $L^2(\ca_{\G})$ and $L^2(\ca_{\G'})$. For example, we require 
\[
P_{\G\G'}\circ D_\G = D_{\G'}\circ P_{\G\G'}\;;
\]
\item[\# 3:]
take the projective and inductive limit of the triples $(B_{\G},D_{\G},H_{\G})$ over all graphs in the inductive system of graphs. Hereby we obtain a candidate for a spectral triple over the full space of connections.
\end{itemize}

This programme was first formulated in \cite{Aastrup:2005yk}
 where we analyzed the possibility to construct a spectral triple based on the projective system of piece-wise analytic graphs used in loop quantum gravity. However, we found this project unlikely to succeed due to the large number of constraints coming from the different projections between embedded graphs.  These constraints restrict the Dirac type operator considerably, and for the loop quantum gravity-setup we were unable to solve them.

To overcome these difficulties we have, in a recent series of publications \cite{Aastrup:2008wa,Aastrup:2008wb,Aastrup:2008zk}, adopted a new strategy where we instead consider a more restricted system of graphs. Examples of these restricted systems of graphs are simplicial complexes/triangulations and their barycentric subdivisions, and hyper-lattices and their sub-lattices obtained by subdividing each cell symmetrically. These examples all permit a spectral triple construction with interesting properties.

\section{The construction}
\label{PRO2}

We now give an outline of the construction of the algebra of holonomy loops and the spectral triple. In the following we work with simplicial complexes and their barycentric subdivisions. The construction based on lattices and their symmetric subdivisions is obtained in the same way. \\

\noindent{\large\bf\# 1: A spectral triple associated to a single graph}\\

First, let $\Gamma$ be a finite $d$-dim simplicial complex with oriented edges $\{\epsilon_i\}$ and vertices $\{v_i\}$, where
$$
\e_j:\{0,1\}\rightarrow \{ v_i \}\;.
$$
\begin{wrapfigure}{r}{30mm}
\resizebox{!}{3cm}{%
   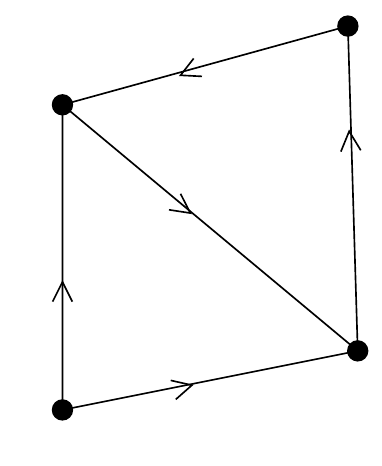}
\end{wrapfigure}
Assign to each edge $\e_i$ a group element $g_i\in G$
\[
\nabla:\e_i\rightarrow g_i\;,
\]
where $G$ is a compact Lie-group. Notice that the space of such maps,  denoted $\ca_\Gamma$, has the form 
$$
\ca_\Gamma\simeq G^n \;,
$$
since
$$
\ca_\Gamma\ni\nabla\rightarrow (\nabla(\e_1),\ldots,\nabla(\e_n))\in G^n\;.
$$

\noindent {\bf The Algebra:} A loop $L$ in $\G$ is a finite oriented sequence of connected edges, $L=\{\e_{i_1},\e_{i_2},\ldots,\e_{i_n}\}$, with $\e_{i_1}(0)=\e_{i_n}(1)$. We choose a fixed basepoint $v_0\in \{v_i\}$ and discard trivial backtracking of loops. 
The product between two loops is given by gluing them at the basepoint
$$
L_1\circ L_2 = \{L_1,L_2\}\;,
$$
and the inversion of a loop is given by reversal
$$
L^*=\{\e_{i_n}^*,\ldots,\e_{i_j}^*,\ldots,\e_{i_1}^*\}\;,
$$
where
$$
\e_j^*(\t) = \e_j(1-\t)\;,\quad \t\in\{0,1\}\;.
$$
Consider formal, finite series of loops
\[
a =\sum_i a_i L_i\;,\quad a_i\in \mathbb{C}\;.
\]
The product between two elements $a$ and $b$ is defined 
\[
a\circ b = \sum_{i,j}(a_i\cdot b_j) L_i\circ L_j\;.
\]
The involution of $a$ is defined
\[
a^* =\sum_i \bar{a}_i L^*_i\;.
\]
These elements have a natural norm
\[
\| a \|= \sup_{\nabla\in\ca_\Gamma}\|\sum a_i \nabla(L_i)  \|_G\;,
\]
where the norm on the rhs is the matrix norm in $G$. Let $B_\Gamma$ be the $C^*$-algebra generated by the $L_i$'s in this norm and let $\cb_\Gamma$ be the $*$-sub-algebra of $B_\Gamma$ generated by the $L_i$'s. We will call this algebra the loop algebra.  \\

\noindent{\bf The Hilbert space:} 
The next step is to consider representations of the loop algebra. In fact, there is a natural Hilbert space
\begin{equation}
\ch_\Gamma= L^2(G^{n},Cl(T^*G^{n})\otimes M_l({C} ))\;,
\label{Hilbert}
\end{equation}
which involves the Clifford bundle over $G^n$. Here $L^2$ is with respect to the
Haar measure on $G^n$ and $l$ is again the size of the representation of $G$.
The reason for adding these two factors -- the Clifford bundle and the matrix factor -- in (\ref{Hilbert}) is to accommodate  the Dirac operator and the algebra of holonomy loops, respectively.

The loop algebra $\cb_\Gamma$ has a natural representation on $\ch_\Gamma$ given by
\[
h_L\cdot \psi(\nabla)= (\One\otimes\nabla(L))\cdot\psi(\nabla)\;,\quad \psi\in H\;,
\]
where the first factor acts on the Clifford-part of the Hilbert space and the second factor acts by matrix multiplication on the matrix part of the Hilbert space. 
For example, the action of the loop

\begin{wrapfigure}{r}{30mm}
\resizebox{!}{4cm}{%
   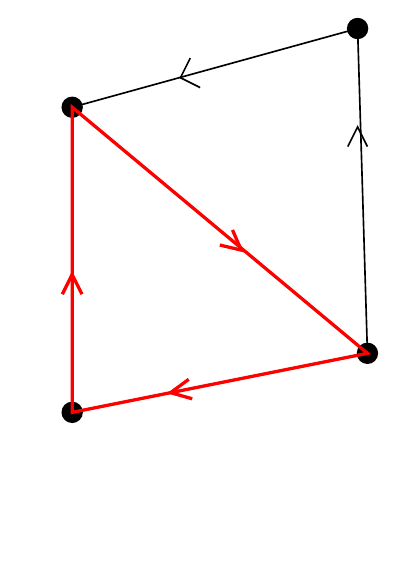}
\end{wrapfigure}
\[
L=\{\e_2,\e_3,\e_1^*\}\;,
\]
which runs in the graphs on the rhs, corresponds to multiplication with the factor 
\[
h_L\sim g_2\cdot g_3 \cdot (g_1)^{-1}
\]
on elements in $\ch_\Gamma$.\\

\noindent{\bf The Dirac operator:} Since $G^n$ is a classical geometry we can simply pick a Dirac operator $D_\Gamma$ on $G^{n}$ 
and hereby obtain 
a candidate for a spectral triple
\[
(\cb_\Gamma,\ch_\Gamma,D_\Gamma)\;,
\]
on the level of the graph $\Gamma$. The choice of the Dirac operator is, at this point, largely free. \\

\newpage
\noindent{\large\bf\# 2: Compatibility with the structure maps}\\

Consider now a system of nested simplicial complexes
\[
\Gamma_0 \rightarrow \Gamma_1 \rightarrow \Gamma_2 \rightarrow \ldots\;,
\]
where $\G_i$ is the barycentric subdivision of $\G_{i-1}$.
For the corresponding spaces $\ca_{\G_i}$ we obtain a system 
\[
 \ca_{\G_0}\;\stackrel{P_{10}}{\longleftarrow} \;\ca_{\G_1}\stackrel{P_{21}}{\longleftarrow} \;\ca_{\G_2}\stackrel{P_{32}}{\longleftarrow}\;\ldots\;,
\]
where the arrows are projections between the spaces $\ca_{\G_i}\simeq G^{n(\G_i)}$. The aim is to obtain a system of spectral triples
\begin{equation}
(\cb_{\Gamma_0},H_{\Gamma_0},D_{\Gamma_0}) \leftrightarrow (\cb_{\Gamma_1},H_{\Gamma_1}  ,D_{\Gamma_1} )\leftrightarrow (\cb_{\Gamma_2},H_{\Gamma_2},D_{\Gamma_2}) \leftrightarrow \ldots\;,
\label{sequence}
\end{equation}
where the arrows indicate compatibility with the structure maps. It turns out that the required compatibility is automatic for the algebra, and is easily obtained for the Hilbert space. For the Dirac operator, however, this requirement strongly restricts the form of permissible operators.

\begin{wrapfigure}{l}{40mm}
\resizebox{!}{3cm}{%
   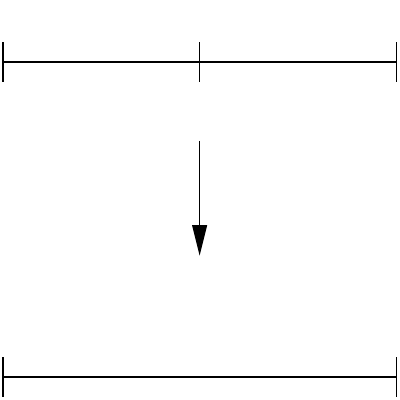}
\end{wrapfigure}
\noindent Essentially, the problem of solving the compatibility conditions boils down to the simple case where we have a single edge which is subdivided into two edges. This corresponds to the projection
\begin{equation}
P: G^2\rightarrow G\; , \quad P(g_1,g_2) = g_1\cdot g_2\;,
\label{proJ}
\end{equation}
and a corresponding map between Hilbert spaces
$$
\hspace{2cm}P^*: L^2(G,Cl(T^*G)\otimes M_l)\rightarrow L^2(G^2,Cl(T^*G)\otimes M_l)\;,
$$
which leads to the compatibility condition
\begin{equation}
P^* (D_1 v)(g_1,g_2) = D_2 (P^*v)(g_1,g_2) \;,\quad v\in L^2(G,Cl(T^*G)\otimes M_l)
\label{condition}
\end{equation}
for the Dirac operator. Here $D_1$ denotes a Dirac operator on $G$,  and $D_2$ denotes a Dirac operator on $G^2$. In the papers \cite{Aastrup:2008wb,Aastrup:2008zk} we have found two different solutions to this compatibility condition. Since the solution described in \cite{Aastrup:2008wb} is considerably more complicated than the solution described in \cite{Aastrup:2008zk} we shall here describe the latter. Consider therefore  the following change of variables
\[
\Theta: G^2\rightarrow G^2\; ;   (g_1,g_2)\rightarrow (g_1\cdot g_2,g_2)=: (g_1',g_2')\;,
\]
which changes the projection (\ref{proJ}) into the simple form
\begin{equation}
P(g_1',g_2') = g_1'\;.
\label{proJ2}
\end{equation}
It is now straight forward to write down a Dirac operator on $G^2$ which is compatible with the projection (\ref{proJ2}). Basically, we can pick any Dirac operator of the form
\begin{equation}
D_2  = D_1 + a D'_2\;,\quad a\in \mathbb{R}\;,
\label{general}
\end{equation}
where $D'_2$ is a Dirac operator on the copy of $G$ in $G^2$ whose coordinates are eliminated by the projection (\ref{proJ2}). At this point the choice of the operator $D'_2$ is essentially unrestricted with $a$ being an arbitrary real parameter. However, for reasons explained in \cite{Aastrup:2008zk} it turns out that  $D_1$ and $D'_2$ should of the form
\begin{equation}
D_i = \sum_{i,a} {\bf e}_i^a \cdot d_{{\bf e}_i^a}\;,
\label{dir}
\end{equation}
where ${\bf e}_i^a$ denotes a complete set of left-invariant vector fields in $TG_i$, $a$ is a $SU(2)$ index, and the product in (\ref{dir}) is Clifford multiplication.

This line of analysis is straightforwardly generalized to repeated subdivisions and gives rise to a series of free parameters $\{a_i\}$, one for each subdivision. 
By solving the $G^2\rightarrow G$ problem repeatedly, and by piecing together the different edges, we end up with a Dirac-like operator on the level of $\Gamma_n$ of the general form
\begin{equation}
D_{\G_n}=\sum_k a_k D_k\;,
\label{general}
\end{equation}
where $D_k$ is a Dirac type operator corresponding to the $k$'th level.
As already mentioned, there are several possible Dirac operators which satisfy condition (\ref{condition}). However, all solutions will be of the general form (\ref{general}) which involves a sequence of free parameters. The solution which we have outlined here and which is published in \cite{Aastrup:2008zk} entails a Dirac type operator with an easily calculable spectrum. \\

\noindent{\large\bf\# 3: The continuum limit}\\

We are now ready to take the limit of the sequence (\ref{sequence}). First, the Hilbert space $\ch_{\smalltriangleup}$ is constructed by adding all the intermediate Hilbert spaces 
$$
\ch' =\oplus_{\G}L^2 (G^{n(\G)},Cl(T^*G^{n(\G)})\otimes M_l({C}))/N\;, 
$$
where $N$ is the subspace generated by elements of the form 
$$
(\ldots , v, \ldots , -P^*_{ij}(v),\ldots )\;,
$$
where $P^*_{ij}$ are the induced maps between Hilbert spaces. The Hilbert space $\ch_{\smalltriangleup}$ is the completion of $\ch'$. The inner product on $\ch_{\smalltriangleup}$ is the inductive limit inner product. This Hilbert space is manifestly separable. 

In fact, the Hilbert space $\ch_{\smalltriangleup}$ is constructed in the same way as the kinematical Hilbert space in loop quantum gravity. If we ignore the Clifford bundle, then the only difference between $\ch_{\smalltriangleup}$ and the kinematical Hilbert space is the choice of graphs, triangulations vs. piece-wise analytic, which corresponds to the separability of $\ch_{\smalltriangleup}$ and the non-separability of the kinematical Hilbert space of loop quantum gravity. We shall discuss this issue in section \ref{comparisson}.

Next, the algebra
$$
\cb_{\smalltriangleup}:= \lim_{\stackrel{\G}{\longrightarrow}}\cb_{\G}\;
$$
contains loops defined on a simplicial complex $\G_n$ in $\{\G_n\}$. Again, the algebra $\cb_{\smalltriangleup}$ is separable.
Finally, the Dirac-like operator $D_{\G_n}$ descends to a densely defined operator on the limit space $\ch_{\smalltriangleup}$
$$
D_{\smalltriangleup} = \lim_{\stackrel{\G}{\longrightarrow}}D_{\G_n}\;.
$$

In \cite{Aastrup:2008wb} we prove\footnote{To be precise, we prove in \cite{Aastrup:2008wb} that a semi-finite spectral triple exist for all compact Lie-groups but that certain deviations to the form described here may be required. For $SU(2)$ these deviations are not necessary.} that for a compact Lie-group $G$ the triple $(\cb_{\smalltriangleup},\ch_{\smalltriangleup},D_{\smalltriangleup})$ is a semi-finite spectral triple, which means that:
\begin{enumerate}
\item
$D_{\smalltriangleup}$'s resolvent  
$
(1+D_{\smalltriangleup}^2)^{-1}
$
is compact (wrt. trace) and
\item
the commutator $[D_{\smalltriangleup},a]$ is bounded,
\end{enumerate}
provided the sequence $\{a_i\}$ approaches $\infty$ sufficiently fast.

The term {\it semi-finite} means, in this case, that the resolvent of $D_{\smalltriangleup}$ is in fact not compact. Instead, the degeneracy of each eigenvalue is finite with respect to a certain trace and the resolvent is compact with respect to this trace. The trace is the ordinary operator trace on operators in $L^2(\overline{\ca}^{\smalltriangleup} )$ tensored with the finite trace on the CAR algebra. The CAR algebra in this setup appears as
\[
\lim_n Cl(T^*_{id} \ca_{\G_n})\;.
\]

For the case $G=U(1)$ we find that the sequence $\{a_n\}$ is required to satisfy
\[
a_n = 2^n b_n\;,\quad \mbox{where}\quad \lim_{n\rightarrow\infty}b_n = \infty \;.
\]
The exact restrictions on the sequence $\{a_n\}$ for arbitrary $G$ may depend on which Dirac-type operator one chooses.

\section{Spaces of connections}
\label{SPCO}

Let us now turn to the spaces $\ca_\G$ and their projective limit.
Denote by
\[
\overline{\ca}^{\smalltriangleup} := \lim_{\stackrel{\G}{\longleftarrow}}\ca_\G\;.
\]
It turns out that the limit space $\overline{\ca}^{\smalltriangleup}$ is a space of generalized connections which means that the space $\ca$ of smooth connections is densely embedded in $\overline{\ca}^{\smalltriangleup}$. This result mirrors the fundamental result of loop quantum gravity mentioned in section \ref{LQG} (see equation (\ref{LQGresult})). 

To see that $\overline{\ca}^{\smalltriangleup}$ contains all smooth connections we first need to map the graphs $\{\Gamma_j\}$ into a manifold $M$
\[
h:\Gamma_j\rightarrow\Gamma_j\in M\;.
\]
Then there is a natural map 
\[
\chi:\ca\rightarrow\overline{\ca}^{\smalltriangleup}\;,\quad\chi(\nabla)(\e_i)=Hol(\nabla,\e_i)\;,
\]
where $Hol(\nabla,\e_i)$ is again the holonomy of $\nabla$ along $\e_i$ which is now in $M$.
In fact, $\chi$ is an embedding
\[
\ca \hookrightarrow  \overline{\ca}^{\smalltriangleup}\;.
\]
\begin{wrapfigure}{l}{40mm}
\resizebox{!}{3.5cm}{%
   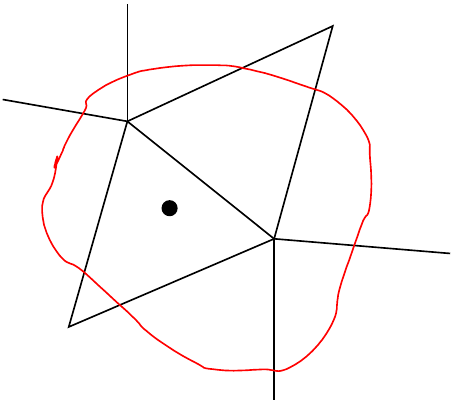}
\end{wrapfigure}
To see this consider two different connections $\nabla_1,\nabla_2\in\ca$ which differ in a point $m\in M$ and therefore in a neighborhood $U$ of $m$. Choose a small edge $\e_i$ in a graphs $\Gamma_j$ so that $\e_i\in U$. Then 
 $$
 Hol(\nabla_1,\e_i)\not=Hol(\nabla_2,\e_i)\;.
 $$ 
 This shows exactly that $\overline{\ca}^{\smalltriangleup}$ contains all smooth connections. Furthermore, the image of $\chi$ is dense in $\overline{\ca}^{\smalltriangleup}$, see \cite{Aastrup:2008wb}. These result suggests that the Dirac type operator is a kind of functional derivation operator over $\ca$
\[
D_{\smalltriangleup}\sim \frac{\d}{\d\nabla}\;.
\]
In fact, $D_{\smalltriangleup}$ is a global operator of the heuristic form
\begin{equation}
D_{\smalltriangleup}\sim \sum_{x} \overline{\nabla(x)}\cdot\frac{\d}{\d\nabla(x)}\;,
\label{tiktak}
\end{equation}
where $\overline{\nabla(x)}$ represents a degree of freedom in each point.

Furthermore, the inner product of the Hilbert space is a functional integral over $\ca$
\[
\langle \Psi | ...|\Psi\rangle \sim\int_{\overline{\ca}^{\smalltriangleup}} [d\nabla] \mbox{Tr}\ldots\;.
\]
In total, these observations suggest that the spectral triple construction should be interpreted in terms of quantum field theory.

\subsection{Comparison to loop quantum gravity}
\label{comparisson}

As already mentioned, loop quantum gravity operates with the space $\overline{\ca}^a$ of generalized connections based on a projective system of piecewise analytic graphs.
Thus, there are now two different completions\footnote{In fact, there are many completions of $\ca$, one for each set of graphs which satisfy certain 'density' conditions \cite{Aastrup:2008wb}.} of the space $\ca$ of smooth connections: $\overline{\ca}^a$ and $\overline{\ca}^{\smalltriangleup}$ with
$$
\ca \hookrightarrow  \overline{\ca}^a\;,\quad \ca \hookrightarrow  \overline{\ca}^{\smalltriangleup}\;.
$$
The difference between these completions is their corresponding symmetry groups:
\begin{enumerate}
\item
In loop quantum gravity the group of analytic diffeomorphisms act as a symmetry group,
\item 
In the present case we have a much smaller symmetry group of graph automorphisms which we denote by Diff$(\triangle)$. 
\end{enumerate}
We observe the following hierachy:
\[
\begin{array}{ll}
\ca: & \mbox{- action of $\mbox{diff}(\cm)$} \\
& \mbox{- no Hilbert space structure}\\
& \mbox{- no Dirac-like operator}\vspace{5mm}\\
\overline{\ca}^a: & \mbox{- action of (analytic) $\mbox{diff}(\cm)$} \\
& \mbox{- Hilbert space structure (non-separable)}\\
& \mbox{- no Dirac-like operator}\vspace{5mm}\\
\overline{\ca}^{\smalltriangleup}: & \mbox{- no action of $\mbox{diff}(\cm)$ (few discrete)}\\
& \mbox{- Hilbert space structure (separable)}\\
& \mbox{- Dirac-like operator}
\end{array}
\]
It appears that the use of a restricted system of graphs (simplicial complexes, cubic lattices or something else) corresponds to a kind of (partial) gauge fixing of the diffeomorphism group.
There is, however, an alternative interpretation. Notice that a triangulation $\Gamma$ is also a piecewise analytic graph\footnote{If we choose the triangulation to be piece-wise analytic.}. This means that there is a natural embedding between the Hilbert spaces
$$
 L^2(\overline{\ca}^{\smalltriangleup}) \stackrel{\iota}{\hookrightarrow} L^2(\overline{\ca}^a)\;,
$$
where we for now discard the Clifford bundle and the matrix factor from $\ch_{\smalltriangleup}$.
Furthermore, in loop quantum gravity there is the Hilbert space $H_{diff}$ of (spatial)  diffeomorphism invariant states. That is, there exist a surjection
\[
L^2(\overline{\ca}^a)\stackrel{q}{\to} \ch_{diff} \;.
\]
We therefore get a map
$${
 L^2(\overline{\ca}^{\smalltriangleup})\stackrel{\Xi}{\longrightarrow} \ch_{diff}\;,
}$$
and obtain the diagram:
\begin{eqnarray}
{
\begin{array}{ccc}
&  &   L^2(\overline{\ca}^a)\\
&\stackrel{\iota}{\nearrow} &\downarrow q\\
 L^2(\overline{\ca}^{\smalltriangleup}) &\stackrel{\Xi}{\longrightarrow} &\ch_{diff}
\end{array}
\nonumber
}
\end{eqnarray}
The amount by which the map $\Xi$ fails to be injective can be thought of as a definition of the symmetry group {diff$(\triangle)$} of graph automorphisms.

This suggests that $\ch_{\smalltriangleup}$ is directly related to the Hilbert space of (spatial) diffeomorphism invariant states known from loop quantum gravity. 
In this picture we can therefore think of a holonomy loop in $B_{\smalltriangleup}$ as an equivalence class of holonomy loops, up to diffeomorphisms.

There are several unresolved issues here. For example, it seems that there is an important difference between the system of cubic lattices and the system of triangulations: The valence number of vertices in the first case is always fixed, determined by the dimension of the manifold $M$, whereas the valence of vertices in the case of triangulations diverges. Thus, for the map $\Xi$ to be surjective we probably must require the valence of vertices to approach infinity. The exact relation between the Hilbert spaces $H_{\smalltriangleup}$ and $H_{diff}$ remains to be analyzed.

\section{The Poisson structure of general relativity}
\label{POISSON}

The aim of this section is to show that the spectral triple construction quantizes the Poisson bracket (\ref{bracket}). To see this we first calculate the commutator between the Dirac type operator $D_{\smalltriangleup}$ and an element of the algebra $B_{\smalltriangleup}$. 
Take first a single group element $g$ corresponding to the $i'th$ copy of $G$ in $G^n$. We find (for simplicity we set $a_i=1$)
\begin{equation}
[D_{\smalltriangleup},g]=\sum_{k}\big(\pm g\sigma_k  \big)\cdot {\bf e}_k \;,
\label{comm}
\end{equation}
 where ${\bf e}_k \in Cl(T^\ast G^n)$ and $\sigma_k$ is a certain generator of the Lie algebra $\mathfrak{g}$.
Next, the commutator between $D$ and the element $h_L\sim g_{i_1}\cdot g_{i_2}\ldots g_{i_k} $ is
\begin{equation}
[D_{\smalltriangleup},h_L]= [D_{\smalltriangleup},g_{i_1}]g_{i_2}\ldots g_{i_k} + g_{i_1}[D_{\smalltriangleup},g_{i_2}]\ldots g_{i_k} + \ldots\;,
\label{comm2}
\end{equation}
which shows that the action of $D_{\smalltriangleup}$ is to insert Lie-algebra generators at each vertex in the loop. Already here we notice a resemblance to the bracket (\ref{bracket}) which also prescribes insertions of Lie-algebra generators into the holonomy loop.

In fact, if we omit the part which involves Clifford multiplication with the elements ${\bf e}_k \in Cl(T^\ast G^n)$ then the bracket given by (\ref{comm}) and (\ref{comm2}) is essentially a quantized version of the Poisson bracket (\ref{bracket}). Here the holonomy loops belong to $B_{\smalltriangleup}$ and the corresponding flux operators, which are now located at the vertices in the graphs $\{\G_i\}$, corresponds to the left-invariant vector fields used to construct the Dirac type operator $D_{\smalltriangleup}$. These left-invariant vector fields should be interpreted as flux operators corresponding to {\it infinitesimal surfaces} located at the endpoints of the edges carrying the copies of $G$ on which the vector fields act. This implies that the Dirac type operator $D_{\smalltriangleup}$ can be written as a sum of {\it all} the flux operators located at the vertices in the family of graphs $\{\G_i\}$.

This shows that the spectral triple construction captures information tantamount to a representation of the Poisson brackets of general relativity. The combination of the algebra of holonomy loops and the Dirac type operator involves both conjugate variables. The main difference to the representation used in loop quantum gravity is the choice of graphs. It is important to keep in mind that the system of graphs which we use here is dense in a double sense: the set of vertices $\{v_i\}$ is a dense set in $M$ and the space of smooth connections is densely embedded in $\overline{\ca}^{\smalltriangleup}$.

If we interpret the Hilbert space $H_{\smalltriangleup}$ in terms of a partial solution to the (spatial) diffeomorphism constraint, then we can think of our construction as a quantization scheme which deals {\bf first} with the constraints (partially) and {\bf next} with the quantization. That is, a quantization scheme which deviates from the standard Dirac-type quantization procedure.

\section{The square of $D_{\smalltriangleup}$ }

In loop quantum gravity the {\bf area operators} play an important role. Given a surface $S$ in $M$ the associated area operator reads \cite{Ashtekar:1996eg}
\[
{\bf A}(S)=  \sum_{n} \sqrt{\hat{ F}^i_{S_n}\hat{ F}^j_{S_n}\d_{ij}}\;.
\]
where $S=\bigcup_n S_n$. Since the area operators are functionals of the flux operators, and since we have just argued that the Dirac type operator $D_{\smalltriangleup}$ has the form of a sum of flux operators, it is natural to ask which role area operators plays in the spectral triple construction discussed here. What we find is that the square of the Dirac type operator $D_{\smalltriangleup}$ has the form of a kind of global area operator. To see this it suffices to note that the square of $D_{\smalltriangleup}$ will, to leading order, be a sum of flux operators squared.
 Thus, we find that
\begin{equation}
D^2_{\smalltriangleup}\sim\sum_v \ldots {\bf A}(S_v)\sim \int_\Sigma  {\bf A}(x)\;,
\label{D2}
\end{equation}
where the sum runs over vertices $\{v_i\}$ and where ${\bf A}(x)$ is a kind of area density squared operator. The integral over $\S$ in (\ref{D2}) should be understood in the sense that the sum $\sum_v$ runs over all vertices in $\{\G_i\}$ which is a dense set in $M$.

It remains to understand exactly which classical interpretation the operator $D_{\smalltriangleup}^2$ should be given. \\

\noindent{\bf The spectral action functional}\\

Let us here also mention that the spectral action functional -- the operator trace over the heat-kernel -- resembles a Feynman integral
\[
Tr \exp(-s (D_{\smalltriangleup})^2)\sim\int_{\overline{\ca}^{\smalltriangleup}}[d\nabla]\exp\left(-s( D_{\smalltriangleup})^2\right)\d_{\nabla}(\nabla)\;,
\]
where $D^2_{\smalltriangleup}$ plays the role of a kind of action or an energy. This object is, for suitable sequences $\{a_n\}$, finite. Notice also that this object is well defined for any compact group $G$ and any dimension of the underlying manifold $M$. It is an interesting question if this bears any relation to ordinary quantum field theory.

\section{The Hamiltonian}
\label{HAM}

A key motivation for constructing the operator $D_{\smalltriangleup}$ is to find canonical structures at a quantized level. The spectral action construction represents a top-down approach to quantum gravity. This means that the relevant question is not "how to quantize" certain structures but rather how to obtain a semi-classical analysis once the construction is completed.

Thus, a relevant question is whether the spectral triple construction has anything to say about the dynamics of quantum gravity? Does it contain information about the Hamiltonian of general relativity?
In the following we will attempt to answer this question with an argument as to how a candidate for a Hamilton constraint might arise from the spectral triple construction.

We start with a fluctuation of the Dirac operator. This is generally of the form 
\[
\tilde{D}_{\smalltriangleup} =D_{\smalltriangleup}+W\;,
\label{inner}
\]
where $W=a[D_{\smalltriangleup},b]$ is a noncommutative 1-form and where $a$ and $b$ are elements of the algebra. $W$ parametrizes a freedom of choosing the Dirac type operator $D_{\smalltriangleup}$ and should be interpreted as a kind of gauge potential {\it over} the space $\overline{\ca}^{\smalltriangleup}$. 
Consider now the square of the fluctuated operator
\begin{equation}
\tilde{D}_{\smalltriangleup}^2 = D_{\smalltriangleup} ^2 + (\tilde{D}_{\smalltriangleup}^2-D_{\smalltriangleup} ^2)= D_{\smalltriangleup} ^2 +\cf_W\;,
\label{D2}
\end{equation}
where $\cf_W= \{D_{\smalltriangleup} ,W\} + \frac{1}{2}\{W,W\}$. We now propose that the quantized Hamiltonian constraint should be of the form 
\begin{equation}
\mbox{Tr}(\cf_W\Psi) =0\;.
\label{Qconstraint}
\end{equation}
Here, Tr is with respect to $SU(2)$. 
Heuristically, this expression has the right form. To see this we first expand the loop $L$ according to 
\[
L\sim 1 +\alpha^2 F +\cdots\;,
\]
where $\alpha^2$ is the area expanded by the "small" loop and $F$ is the field strength tensor of the Ashtekar connection.
Furthermore, $D_{\smalltriangleup}$ is heuristically of the form
$$
D_{\smalltriangleup}= \int_\Sigma dx {\bf e}^i_j(x) \cdot\hat{E}_i^j(x)\;,
$$ 
which can be seen from (\ref{tiktak}) and from the fact that the sum runs over a dense set in $M$. Here ${\bf e}^i_j(x) $ represents again an element in the Clifford bundle.
When we combine these expressions we obtain, to second order in $\alpha$, the following:
\begin{eqnarray*}
\lefteqn{\mbox{Tr}(\tilde{D}_{\smalltriangleup}^2-D_{\smalltriangleup}^2)}\\
&
=&
\int_\Sigma \int_\Sigma dxdy\{ {\bf e}^i_j(x) \hat{E}_i^j(x), [{\bf e}^k_l(y) \hat{E}_k^l(y), \alpha^2 F]\}\;.
\end{eqnarray*}
However the anti commutator of $ {\bf e}^i_j(x)$ and ${\bf e}^k_l(y)$ will produce a delta function in $x-y$, and we get an expression similar to  
 \[
 \int dx\epsilon^{ij}_k\hat{F}^k_{\mu \nu}(x)\hat{E}_j^\mu(x) \hat{E}_i^\nu(x)\;,
 \]
which has the same structure as (\ref{HAM}) and therefore resembles a quantization of the Hamilton constraint.

It is important to emphasize that this argument, in its present form, only works at a heuristic level. We expect that much can be added to the argument in the form of additional structure. For instance, when discussing fluctuation of the Dirac operator one should also consider a real structure, which, at present, has not been analyzed nor constructed. Furthermore, a rigorous semiclassical analysis is necessary to fully estimate the significance of an expression like  (\ref{Qconstraint}).
 
 However, 
 it is clear that once the spectral triple $(\cb_{\smalltriangleup},\ch_{\smalltriangleup},D_{\smalltriangleup})$ is constructed, then an object like expression (\ref{D2}) and the constraint (\ref{Qconstraint}) are canonical and free of the ordering ambiguities which one encounters in a standard quantization scheme. Also, these expression require no regularization. This indicates that a spectral triple construction like the one discussed here might indeed help define a canonical theory of quantum gravity.


\section{The sequence $\{a_i\}$}

\begin{wrapfigure}{r}{45mm}
\resizebox{!}{3cm}{%
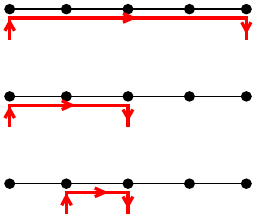 }
\end{wrapfigure}

Recall that the Dirac operator $D_{\smalltriangleup}$ depends on a sequence $\{a_i\}$ of real parameters. To understand and interpret the spectral triple construction it is necessary to understand the significance of these parameters. 

First, the role of the parameters $\{a_i\}$ is to set a scale. Edges which lie 'deep' in the inductive system of graphs are assigned large $a$'s compared to edges which lie 'higher' (see figure). This means that
a 'coarse grained' loop corresponds to small $a$'s whereas 
a 'refined' loop corresponds to large $a$'s.

Therefore, when $D_{\smalltriangleup}$ probes a loop the eigenvalues corresponding to 'coarse' information about the loop are weighted with smaller $a$'s compared to eigenvalues which corresponds to more 'refined' information. 

It is the requirement of $D_{\smalltriangleup}$ to have compact resolvent which dictates this additional scaling degree of freedom. At present we have no clear interpretation of this structure, except that this seems important in order to extract topological information of the underlying manifold. Furthermore, such a scaling degree of freedom is reminiscent of the renormalization group theory of lattice gauge theory.

There is in fact a natural choice for the $a$'s since they correspond to a split of an edge in two. This suggest the sequence 
\[
a_n=2^{n}\;.
\]
However, although this sequence is divergent, it is in fact exactly the sequence where $D_{\smalltriangleup}$ fails to be spectral. This means that infinities (for example, in the spectral action functional) will arise for exactly this choice\footnote{This statement probably depends on the exact choice of Dirac type operator. We know it is valid for the Dirac type operators constructed in \cite{Aastrup:2008zk}.}.
Instead one could try with the permissible sequence 
\[
a_n=(2+\epsilon)^{n}\;,
\]
and subsequently take the limit $\epsilon\rightarrow 0$. One may speculate whether the infinities which arise in this limit might somehow be related to the infinities one encounters in the renormalization procedure of quantum field theory.

\section{Connes Distance Formula}

Given a spectral triple $(A,D,H)$ over a manifold $M$ Connes spectral distance formula reads
\[
d(\xi_x,\xi_y)=\sup_{a\in\ca}\big\{ |\xi_x(a)-\xi_y(a)| \big|  |[D,a]|\leq 1 \big\}\;,
\]
where $\xi_x,\xi_y$ are the evaluation homomorphisms $A\rightarrow\mathbb{C}$ in $x,y$ and where $d(\xi_x,\xi_y)$ is the geodesic distance between $x$ and $y$ on $M$ given by metric structure of $D$. The interesting observation, which is central to noncommutative geometry, is that this distance formula can be straight forwardly  generalized to noncommutative spaces and algebras.

For us, the question therefore arises what Connes distance formula has to say about the spectral triple construction which we discuss here. 
It turns out that a distance formula which involves the algebra $B_{\smalltriangleup}$ and the Dirac type operator $D_{\smalltriangleup}$ will provide us with a notion of distance on a space of field configurations. The exact meaning of such a formula depends on the state-space of the algebra $B_{\smalltriangleup}$. Without going into detail let us here just state that two field configurations, each given by a connection, will be 'far' apart if they differ on a large scale and 'close' if they differ by a short scale. The reason for this is again the sequence $\{a_n\}$ of scaling parameters. If the two field configurations differ on edges which corresponds to small $a$'s -- corresponding to large scales -- then the corresponding distance will be relatively smaller compared to edges which are assigned large $a$'s.

\section{Conclussion}

In this note we have outlined the motivation, construction and interpretation of the spectral triple construction already presented in  \cite{Aastrup:2008wa}\cite{Aastrup:2008wb}\cite{Aastrup:2008zk}. To recapitulate, the semi-finite spectral triple $(B_{\smalltriangleup},D_{\smalltriangleup},H_{\smalltriangleup})$ consist of:
\begin{itemize}
\item[-]
the algebra $B_{\smalltriangleup}$ of holonomy loops,
\item[-]
the Dirac type operator $D_{\smalltriangleup}$ which resembles a global functional derivation operator,
\item[-]
the Hilbert space $H_{\smalltriangleup}$ which is related to the Hilbert space of (spatial) diffeomorphism invariant states.
\end{itemize}
Furthermore, the interaction between $B_{\smalltriangleup}$ and $D_{\smalltriangleup}$ encodes the quantized Poisson structure of general relativity when formulated in terms of Ashtekar variables. 

Technically, the triple is constructed as a projective/inductive limit of finite dimensional spectral triples associated to oriented graphs. Also, the construction of the spectral triple relies on a set $\{a_i\}$ of scaling parameters. It is the correct scaling behavior of these parameters which ensures that the construction meets the requirements of a spectral triple.

Finally, we present an argument as to how a quantized Hamiltonian of general relativity might emerge from a spectral triple construction like the one presented here. This heuristic argument provides a new, top-down approach to the formulation of a dynamical principle of quantum gravity.\\

\noindent{\bf\large Acknowledgements}\\

We are thankful to Mario Paschke for numerous fruitful discussions on the Hamilton constraint. Also, we thank Christian Fleischhack for very helpful discussions.

\end{document}